\documentstyle[11pt]{article}
\newlength{\bredde}
\def\slash#1{\settowidth{\bredde}{$#1$}\ifmmode\,\raisebox{.15ex}{/}
\hspace*{-\bredde} #1\else$\,\raisebox{.15ex}{/}\hspace*{-\bredde} #1$\fi}
\newcommand{\beq}{\begin{equation}}
\newcommand{\eeq}{\end{equation}}
\newcommand{\bea}{\begin{eqnarray}}
\newcommand{\eea}{\end{eqnarray}}

\def\gtwid{\raise.3ex\hbox{$>$\kern-.75em\lower1ex\hbox{$\sim$}}}
\def\ltwid{\raise.3ex\hbox{$<$\kern-.75em\lower1ex\hbox{$\sim$}}}
\newcommand{\dirac}[1]{/ \!\!\! #1}
\newcommand{\Dt}{/ \!\!\!\! D_\theta}
\newcommand{\gv}{\gamma_5}

\newsavebox{\uuunit}
\sbox{\uuunit}
    {\setlength{\unitlength}{0.825em}
     \begin{picture}(0.6,0.7)
        \thinlines
        \put(0,0){\line(1,0){0.5}}
        \put(0.15,0){\line(0,1){0.7}}
        \put(0.35,0){\line(0,1){0.8}}
       \multiput(0.3,0.8)(-0.04,-0.02){12}{\rule{0.5pt}{0.5pt}}
     \end {picture}}
\newcommand {\unity}{\mathord{\!\usebox{\uuunit}}}
\newcommand{\dr}{\delta^r}
\newcommand{\dl}{\delta^l}
\newcommand{\gh}[1]{\mbox{gh} \left( #1 \right)}

\begin{document}
\begin{titlepage}
\title{\Large{
BRST Gauge Fixing and Regularization}}
\vspace{0.5cm}
\author{{\sc P.H. Damgaard}\thanks{On leave of absence from the
Niels Bohr Institute, Blegdamsvej 17, DK-2100 Copenhagen, Denmark.}
\\
Institute of Theoretical Physics, Uppsala University,
S-751 08 Uppsala, Sweden\\
{\sc F. De Jonghe} \\
NIKHEF-H, Postbus  41882, 1009 DB Amsterdam, The Netherlands \\
and \\
{\sc R. Sollacher} \\
Institute of Theoretical Physics, Uppsala University,
S-751 08 Uppsala, Sweden
}

\maketitle
\begin{abstract} In the presence of consistent regulators, the standard
procedure of BRST gauge fixing (or moving from one gauge to another)
can require non-trivial modifications. These modifications occur at
the quantum level, and gauges exist which are only well-defined when
quantum mechanical modifications are correctly taken into account.
We illustrate how this phenomenon manifests itself in the solvable
case of two-dimensional bosonization in the path-integral
formalism. As a by-product, we show how to derive smooth bosonization
in Batalin-Vilkovisky Lagrangian BRST quantization.
\end{abstract}
\vfill
\begin{flushright}
UUITP-8/95\\
NIKHEF-H 95-017\\
hep-th/9505040

\end{flushright}

\end{titlepage}
\newpage

{\bf 1.}
Essentially all approaches to gauge fixing in the path integral
formalism are based on formal manipulations that ignore the issue
of regularization. It is known, moreover, that care may be required
if one is to work in the framework of a (perturbatively) regularized
quantum field theory.
For a recent discussion of this issue in the
context of Yang-Mills theory, see ref. \cite{Dejonghe}.

Actually, the problem of a suitable BRST gauge-fixing procedure in the
{\em regularized} path integral goes even deeper. As we shall describe
in this letter, there are a number of situations in
regularized Lagrangian quantum field theory where at least three
known methods of gauge fixing (Faddeev-Popov \cite{Faddeev}, usual
Lagrangian BRST \cite{who?}, and Batalin-Vilkovisky \cite{Batalin})
all fail to produce a correctly gauge-fixed path integral, when applied
naively. We shall also show how solutions to this problem can be found.

Consider a gauge
theory based on an action $S[\phi]$, and a local gauge transformation
$\delta\phi^i = R^i_\alpha[\phi]\epsilon^\alpha$. Suppose that under this
gauge transformation a certain operator $\cal{O}$ remains invariant at
the classical level, but is modified at the quantum level. Obviously,
at the classical level the operator $\cal{O}$ will be unsuitable for
a gauge-fixing function. Since gauge
fixing refers not just to the classical aspects of the field theory but
also to the quantum level, a natural question to ask is the following:
can one gauge-fix on the operator $\cal{O}$ in the {\em full} theory,
which incorporates quantum effects? This problem is not nearly as
contrived as it may appear at first sight. It arises in the
2-dimensional context of ``smooth bosonization''
if one follows the path of ref.\cite{us1} and in the
general discussion of an effective $U(1)$ flavour-singlet Lagrangian
using collective fields \cite{us2}. In both of these examples, one
essentially has to ``gauge fix on the chiral anomaly'', and for this
reason the solution to that particular gauge-fixing problem was called
``anomalous gauge fixing''. But the problem appears in related contexts
even when the operator in question is not anomalous, only suffers a
certain (non-anomalous) modification at the quantum level \cite{us3}.
In refs. \cite{us1,us2,us3} various valid solutions to this rather
unusual gauge-fixing problems were presented, but no general procedure
was outlined. We emphasize again that modifications to the naive rules
of (BRST) gauge fixing are required in all of these cases, and that these
modifications are inescapable if one wishes to correctly describe the
quantum mechanical phenomena in those gauges.\footnote{In some special
cases a way to circumvent the problem can be found. See below.} An
example of an
analogous gauge-fixing anomaly within the context of string theory is
discussed in ref. \cite{Niemi}.

Clearly, the issue of an operator being invariant at the classical level,
but undergoing modifications due to quantum corrections, is intimately
tied to the choice of regulator. If we abandon the operator language by
working instead in the Lagrangian path integral formalism, the
``operators''
remain unchanged when manipulated inside the functional integral, but the
analogues of
the quantum corrections show up due to effects of the regulators. It is
here convenient to consider regularization in a (generalized)
Pauli-Villars sense.
At the one-loop level, an
explicit correspondence
between this kind of regularization and a Fujikawa-type \cite{Fujikawa}
path-integral regularization of the {\em measure} can be derived
\cite{Diaz}.
At the one-loop level we can thus consider the whole regularization as
arising from the path integral measure. It is important to keep in mind
that in general consistent regularizations of the path integral measure
will depend non-trivially on the action $S[\phi]$ under
consideration. From the point of view that the
regularized path-integral measure can be obtained at
the one-loop level from a particular Pauli-Villars scheme by integrating
out regularizing ``Pauli-Villars fields'' (using suitable integration
rules \cite{Diaz}), this is of course entirely natural.

If the regularized path integral measure remains invariant under
the local gauge transformation and the associated global BRST
transformation,
then no problems of the kind discussed above can arise. This is obvious,
since (at least to the one-loop level at which we are restricting
ourselves for the moment) no operators $\cal{O}$ that are classically
gauge (or BRST) invariant do not remain invariant at the quantum level.
It does not mean, however, that gauge fixing
-- or in a more precise language: once fixed, going
from one gauge to another -- is an entirely trivial matter. In the usual
Lagrangian BRST formalism going from one gauge to another can be achieved
by adding a ``gauge fermion'' of the type $\delta \Psi[\phi]$ (where
$\delta$ here refers to the associated BRST variation) to the
previous action:
\beq
S[\phi] ~\to~ \tilde{S}[\phi] = S[\phi] + \delta \Psi[\phi] ~.
\eeq
Clearly, adding this new term will in general affect the needed
regularization as well. So the functional measure must be modified
appropriately:
\beq
\int[d\phi]_S \exp\left[\frac{i}{\hbar}S[\phi]\right] ~\to~
\int[d\phi]_{\tilde{S}} \exp\left[\frac{i}{\hbar}\{\tilde{S}[\phi] +
\delta \Psi[\phi]\}\right]~,
\eeq
where $\phi^A(x)$ stands for the generic collection of classical fields
$\phi^i$, ghosts, antighosts, auxiliary fields, ghosts-for-ghosts, etc.,
as needed. This required modification of the functional measure
simultaneously with adding gauge fermions is the source of new
(and often surprising) effects that may arise when the gauge-fixing
operators transform differently at the classical and quantum level.

Consider now the possibility that neither the action nor the measure,
but only the combination of the two, is invariant under the BRST
transformation. To the one-loop level this means that the partition
function must be of the form
\beq
Z = \int [d\phi]_{S} \exp\left[\frac{i}{\hbar}\{S[\phi]
+ \hbar {\cal M}_1[\phi]\}\right] ~,
\label{Zreg1}
\eeq
where the BRST variation $i\delta {\cal M}_1$ precisely cancels the
logarithm of the
Jacobian. Non-invariance of the measure can arise either as a result of
the regularization, or be inherent to the formalism, as in, $e.g.$, the
case of open gauge algebras. There may or may not be higher order
contributions (in $\hbar$) to the action, and we shall return to this more
general all-order situation below, but for the present a one-loop
consideration will suffice. Notice that the measure is regularized only
with respect to $S$ (and not $S + \hbar {\cal M}_1$), since either
${\cal M}_1$ does not
affect the regularization, or, if it does, the effect of modifying the
measure on account of this will be of higher order in $\hbar$. What
happens now when we add the BRST-variation of a gauge fermion?
Formally, i.e.,
when ignoring issues of regularisation, adding a term of the form
$\delta \Psi[\phi]$ to the action does not affect BRST invariance of
the functional integral (since $\delta^2 = 0$), and does not affect the
expectation values of BRST-invariant observables. Let us try the same
procedure in the regularized path integral (\ref{Zreg1}):
\beq
\int [d\phi]_{S} \exp\left[\frac{i}{\hbar}\{S[\phi]
+ \hbar {\cal M}_1[\phi]\}\right] ~\to~ \int [d\phi]_{S+\delta \Psi}
\exp\left[\frac{i}{\hbar}\{S[\phi] + \hbar {\cal M}_1[\phi]+
\delta \Psi[\phi]\}\right ] ~.
\label{Zreg2}
\eeq
The r.h.s. of (\ref{Zreg2}) does not in general describe a BRST-symmetric
path integral. This is due to the fact that the cancellation between
the measure
and $i{\cal M}_1$ in general will be destroyed, since the Jacobian from the
measure is $\delta \Psi$-dependent, while ${\cal M}_1$ is not. So the first
lesson we learn is that {\em in the functional integral, when
regularized through its measure,
adding BRST-exact terms to the action may break the BRST
symmetry}.\footnote{If one insists that the path integral shall remain
properly regularized after the addition of the BRST-exact term to the
action. One can keep BRST symmetry at the cost of losing the proper
regularization, but we discard this possibility.}

At the one-loop level an obvious cure to this problem presents itself.
When we add (or change) a gauge fermion through a term $\delta\Psi$,
we can correct the non-invariance of the measure and action by
re-calculating the one-loop counterterm $i\hbar{\cal M}_1$ by using the
new regularized measure. The final result will then by construction
be BRST invariant to the one-loop level. The only unpleasant by-product
of this procedure is that we seem to lose some of the power of gauge
fixing. Knowing which gauge we are actually going to by adding a term
$\delta\Psi$ will now require a one-loop calculation, and cannot
be read off immediately from the original action.
An alternative route can be found by keeping the regulator fields
in the action throughout, and by requiring that the gauge fermion
$\Psi$ be replaced by a properly regularized $\Psi_R$. The whole
procedure is then again by construction BRST invariant {\em and}
properly regularized. As we shall illustrate below, it will however,
in general lead to rather unusual gauge-fixings in terms of the
original fields. It will also typically lead to
higher-order terms in the ghost fields, once the regulator fields
are integrated out of the path integral.

{\bf 2.}
Let us illustrate these last considerations by the example of
a two-dimensional field theory of Dirac fermions coupled to external
Abelian axial sources
\cite{us1}. The Pauli-Villars--regularized version reads
\begin{eqnarray}  \label{start}
Z[A] &=& \int d[\bar{\psi}, \psi] \prod_l d[\bar{\psi}_l,\psi_l] \exp
\left( \frac{i}{\hbar} S_0 \right)\cr
S_0 &=& \int d^2x \left[ \bar{\psi}i\slash{D}\psi + \sum_l \bar{\psi}_l
(i\slash{D}-M_l) \psi_l \right]
\end{eqnarray}
with $D_\mu = \partial_\mu -i A_\mu \gamma_5$. We choose a
regularization scheme such that the following relations hold (cf. ref.
\cite{Ball}):
\beq
\sum_l c_l =1\quad , \quad \sum_l c_l k_l^m = 0 ~~ {\mbox{\rm for}}~
 m=1,2\quad
, \quad M_l = k_l \Lambda
\label{PVdef}
\eeq
The regulator masses are proportional to a cutoff $\Lambda$ which may
eventually be sent to infinity. For simplicity we take the coefficients
$c_l$ to be $+1$ or $-1$ depending on whether the corresponding
regulator fields are Grassmann-even or -odd.
Although this may look
like a free theory, the coupling to the sources makes it non-trivial,
and by e.g. integrating over these sources with appropriate measures,
we can easily turn it into a genuinely interacting theory. For this
reason we need to regularize the path integral by a suitable number
of Pauli-Villars fields, here labelled by $l$. The measures of all the
fields are otherwise unconstrained,
because the path integral regularization is completely taken
into account (up to one loop order) by the regulator fields $\psi_l$.
This system can conveniently be viewed as an Abelian gauge theory in
disguise. The missing gauge degree of freedom can be re-introduced
through a field redefinitions of the original fields
$\bar{\psi},\psi$. We follow the example of \cite{us1} and choose a
chiral transformation
\begin{equation}
\bar{\psi} = \bar{\chi} e^{i \gamma_5 \theta}\quad , \quad \psi = e^{i
\gamma_5 \theta} \chi ~,
\label{trans}
\end{equation}
which introduces a new pseudoscalar field $\theta (x)$. Because we
regularize the theory by explicit Pauli-Villars fields, and because we
wish to deal only with properly regularized expressions, we let the
regulator fields transform analogously:
\begin{equation}
   \label{transreg}
\bar{\psi_l} = \bar{\chi_l}' e^{i \gamma_5 \theta}\quad , \quad \psi_l
= e^{i\gamma_5 \theta} \chi_l' ~.
\end{equation}
With this choice, the Jacobian of this chiral transformation equals unity.
If we integrate over the field $\theta(x)$ in the path integral,
the resulting generating functional then takes the form
\begin{eqnarray}
Z_{ext}[A] &=& \int d[\theta] d[\bar{\chi}, \chi] \prod_l
d[\bar{\chi}_l',\chi_l'] \exp \left( \frac{i}{\hbar} S_{ext} \right)\cr
S_{ext} &=& \int d^2x \left[ \bar{\chi}i\slash{D}_\theta \chi + \sum_l
\bar{\chi}_l' (i\slash{D}_\theta - M_l^\theta) \chi_l' \right]
\label{Zext}
\end{eqnarray}
where $\slash{D}_\theta = \slash{D} +i\slash{\partial} \theta
\gamma_5$, and $M_l^\theta = M_l \exp (2i\gamma_5 \theta )$. Because
of the redundant introduction
of a new degree of freedom, the ``collective field" $\theta(x)$,
the theory now has a local gauge symmetry \cite{Alf}.
This gauge symmetry is simply
\begin{equation}
\bar{\chi} \to \bar{\chi} e^{i \gamma_5 \alpha}\quad , \quad \chi \to e^{i
\gamma_5 \alpha} \psi \quad ,\quad \theta \to \theta -\alpha ~,
\label{gsym1}
\end{equation}
supplemented with the analogous transformations of the regulator fields.
Now, however, the regulator fields are not the standard ones of
a Pauli-Villars regularization that corresponds to the Dirac operator
$i\slash{D}_\theta - M_l$. In order to
arrive at the standard regulator expressions one can choose to
redefine the regulator fields according to
\begin{equation}
\bar{\chi}_l' = \bar{\chi}_l \quad , \quad \chi_l' =
(i\slash{D}_\theta - M_l^\theta)^{-1}(i\slash{D}_\theta - M_l) \chi_l
\label{newreg}
\end{equation}
This transformation has a non-trivial Jacobian, but it is a computable
Jacobian since it only involves a ratio of determinants. We can thus
write
\begin{eqnarray}
\label{stap2}
Z_{ext}[A] &=& \int d[\theta] d[\bar{\chi}, \chi] \prod_l
d[\bar{\chi}_l,\chi_l] \exp \left( \frac{i}{\hbar}S + i{\cal M}_1
\right)\cr S &=& \int d^2x \left[ \bar{\chi}i\slash{D}_\theta \chi +
\sum_l \bar{\chi}_l
(i\slash{D}_\theta -M_l) \chi_l \right]\cr
{\cal M}_1 &=& -i \ln \det \prod_l \left( \frac{i\slash{D}_\theta
-M_l}{i\slash{D}_\theta -M_l^\theta} \right)^{c_l}
\end{eqnarray}
Here, $c_l$ are the coefficients introduced in (\ref{PVdef}).
The term $M_1$ is the exponentiated Jacobian. It includes
terms that vanish in the limit of infinite
regulator masses. The gauge symmetry (\ref{gsym1}) is now supplemented
by a modified transformation of the regulator fields, which follows
from (\ref{newreg}):
\begin{eqnarray}
\bar{\chi}_l &\to& \bar{\chi}_l e^{i\alpha \gamma_5}\cr
\chi_l &\to& e^{i \alpha \gamma_5} (i\slash{D}_\theta - M_l^\alpha
)^{-1} (i\slash{D}_\theta - M_l) \chi_l~~.
\label{gsym2}
\end{eqnarray}
Obviously, these transformations compensate for those shifts of
$\theta$ that couple
to the regulator fields. In addition, the Jacobian of this
transformation cancels the modification of $M_1$ due to the shift of
$\theta$.
We are now in a position to establish the BRST-transformation rules of
all fields. As usual, we replace the parameter field $\alpha (x)$ in eqs.
(\ref{gsym1}) and (\ref{gsym2}) -- now considered infinitesimal -- by
the ghost field $c(x)$. In addition, we use the standard
transformation rules for the auxiliary fields:
\bea
\delta \bar{\chi} &=& i\bar{\chi} \gamma_5 c\cr
\delta \chi &=& -i c \gamma_5 \chi\cr
\delta \theta &=& -c\cr
\delta \bar{\chi}_l &=& i\bar{\chi}\gamma_5 c \cr
\delta \chi_l &=& -\left( ic\gamma_5 + (i \slash{D}_\theta -M_l
)^{-1} 2iM_l c\gamma_5 \right)\chi_l\cr
\delta \bar{c} &=& b\cr
\delta c &=& \delta b = 0
\label{BRS}
\eea
A straightforward calculation shows that these transformations are
nilpotent, $i.e.$, $\delta^2 =0$.
We may now fix a gauge by adding the BRST-variation $\delta \Psi_R$ of a
properly regularized gauge fermion $\Psi_R$. To see that
highly unusual gauges can be reached in this manner, let us follow
a simplified version of the example given in ref. \cite{us1}. The most
interesting gauge
is the ``bosonization gauge". In (1+1) dimensions such a gauge can be
found in the present context by eliminating the divergence of the
axial fermionic current.\footnote{This ignores the gauge fixing of
the zero mode corresponding to constants shifts of $\theta$. As we
use the example for illustration only, we shall ignore this subtlety
here. The correct treatment of the zero mode can be found in ref.
\cite{us1}.} Note that we are not trying to gauge-fix the original
fermionic degrees of freedom away (an impossible task, since they
have physical significance), only the
chirally rotated fermionic degrees of freedom. In this way we can trade
fermionic degrees of freedom with bosonic ones, choosing to let the
original physical fermionic currents be described by bosonic
expressions. It is thus
possible to gauge-fix the divergence of the
chirally rotated fermionic axial current to zero. The crucial
incredient is of course that we restrict ourselves to properly
regularized expressions. For that reason we choose the corresponding
gauge fermion $\Psi_R$ to be
\beq
\Psi_R = \bar{c} \partial_\mu \left(\bar{\chi} \gamma^\mu \gamma_5 \chi
+ \sum_l \bar{\chi}_l \gamma^\mu \gamma_5 \chi_l \right)~.
\label{PsiR}
\eeq
With this, the generating functional becomes
\begin{eqnarray}
Z[A] &=& \int d[\theta] d[\bar{\chi}, \chi] \prod_l
d[\bar{\chi}_l,\chi_l] d[b] d[\bar{c},c] \exp \left( \frac{i}{\hbar} S
+ i {\cal M}_1 + i\delta \Psi_R\right)\cr
S &=& \int d^2x \left[ \bar{\chi}i\slash{D}_\theta \chi + \sum_l
\bar{\chi}_l (i\slash{D}_\theta -M_l) \chi_l \right]\cr
{\cal M}_1 &=& -i \ln \det \prod_l \left( \frac{i\slash{D}_\theta
-M_l}{i\slash{D}_\theta -M_l^\theta} \right)^{c_l}\cr
\delta \Psi_R &=& b \partial_\mu \left(\bar{\chi} \gamma^\mu \gamma_5
\chi + \sum_l \bar{\chi}_l \gamma^\mu \gamma_5 \chi_l \right)\cr
&& + \bar{c} \partial_\mu \sum_l \bar{\chi}_l \gamma^\mu \gamma_5
(i\slash{D}_\theta - M_l)^{-1} 2 i M_l \gamma_5 \chi_l c
\label{Zgf}
\end{eqnarray}
Note the crucial r\^{o}le played by the regulator fields in the gauge
fixing proposed in eq. (\ref{PsiR}). The part $\partial_{\mu}\bar{\chi}
\gamma^{\mu}\gamma_5\chi$ is BRST invariant by itself, and as such
of course is not a suitable function to gauge-fix on. Indeed, if we
had just kept that term, the gauge fixing would consist of only
$b\partial_{\mu}\bar{\chi}\gamma^{\mu}\gamma_5\chi$, with no
ghost term. Such a term is also all one would be left with if one
naively applied the formulation in which the regulator fields have been
integrated out from the beginning. Although naively BRST invariant,
it would be incorrect
for a variety of reasons, as will be seen below. The regulator fields are
in eq. (\ref{PsiR}) the only fields responsible for providing the correct
ghost term.
The structure of ${\cal M}_1$ and the ghost term becomes more
enlightened if we apply an expansion in
inverse powers of the cutoff $\Lambda$. For ${\cal M}_1$ this yields
\beq
{\cal M}_1 =
\int\! d^2x\; \left[ \frac{1}{2\pi} \partial_\mu \theta \partial^\mu
\theta -\frac{1}{\pi} A_\mu
\partial^\mu \theta +
\frac{\kappa_{-2}}{12 \pi \Lambda^2} \partial^2
\theta \partial^2 \theta -
\frac{\kappa_{-2}}{6 \pi \Lambda^2} \partial_\nu A_\mu
 \partial_\nu \partial_\mu \theta + {\cal O} (\Lambda^{-4}) \right]
\label{M1}
\eeq
with
\beq
\kappa_{-2} = \sum_l \frac{c_l}{k_l^2}~~.
\eeq
Obviously, ${\cal M}_1$
not only provides a kinetic term for the new
field $\theta$ but also higher derivative terms.
These terms are responsible for a {\em regularized} perturbative
$\theta$-propagator \cite{us1}. In addition,
there are terms with higher power
s of $\partial_\mu \theta$ and the external source $A_{\mu}$;
it is a special feature of
2 dimensions  that these occur only at order ${\cal O} (\Lambda^{-4})$.
Usually, one treates the regularization as part of the measure. This
means that for any step in the abovementioned procedure one has to
adjust the regulator to the fermionic part and integrate out the
regulator fields. For the
ghost term in (\ref{Zgf}) we can do this performing a linked
cluster expansion for the regulator fields.
To leading order in $1/\Lambda$ we get
\beq
\bar{c} \left( -\frac{1}{\pi} \partial^2 + \frac{\kappa_{-2}}
{6\pi\Lambda^2} (\partial^2)^2 +\ldots \right) c + \ldots
\eeq
The remaining terms are all of ${\cal O}(\Lambda^{-2})$. They depend on
higher powers in $A_\mu, \partial_\mu \theta$ and  $\partial_\mu b$.
In addition, there are also terms involving higher powers of the ghost
fields; they are due to higher correlation
functions of $\sum_l \bar{\chi}_l \gamma^\mu \gamma_5
(i\slash{D}_\theta - M_l)^{-1} 2 i M_l \gamma_5 \chi_l $ occuring in the
linked cluster expansion. But they are of higher order in $\hbar$, and
should hence be ignored in this one-loop treatment.
At leading order in $1/\Lambda$ the gauge (\ref{PsiR}) coincides with
the bosonization gauge of ref. \cite{us1}. However, it may also have
become clear that including the regulator fields
implicitly, $i.e.$ as part of the measure, a gauge fixing
procedure for {\em finite} cutoff is far more involved. In particular,
the higher order terms introducing couplings between the ghost themselves,
and between the ghosts and the field $b$, are difficult  to interpret in
standard Faddeev-Popov or BRST language.
Let us now take a closer look at the gauge constraint (\ref{PsiR}).
Obviously, we are removing any occurrences of the divergence of the
regularized fermionic axial current in any correlation function. What
then describes the divergence of the original regularized axial
current $J^{\mu}_5$
in terms of the fields $\bar{\psi}, \psi$ and its regulators?
Recall that
\beq
\langle\partial_{\mu}J^{\mu}_5(x_1) \cdots \partial_{\nu}J^{\nu}_5
(x_n)\rangle = \left[\frac{\delta^n Z[\varphi]}
{\delta\varphi(x_1)\cdot
\delta\varphi(x_n)}\right] ~,
\eeq
where
$A_{\mu} = \partial_{\mu}\varphi + \epsilon_{\mu\nu}\sigma$, and where
$\sigma$ has been removed by a fermionic phase rotation (since we
assume that our regulator preserves vector symmetry).
Shifting $b$ by $\varphi$
decouples the fermions $\bar{\chi},\chi$ and its regulators from
$\varphi$. Functional derivatives with respect to $\varphi$ then get
contributions only from $M_1$ and from the ghost term. This means that
for the present gauge we describe the divergence of the original axial
current by the purely bosonic expression
$-(1/\pi) \partial^2 \theta$, up to terms corresponding
to higher derivatives and powers of $\theta$, suppressed by inverse
powers of the cutoff. So to leading order in $1/\Lambda$, we precisely
get the sought-for identification.
Note that for finite cut-off $\Lambda$ we are left among others with
highly unusual
ghost terms, which could not straightforwardly have been
derived by a Faddeev-Popov method.

{\bf 3.}
We now translate the main steps of the previous section
into the language of
(regularised) Batalin-Vilkovisky (BV) Lagrangian quantisation
\cite{Batalin}. Since no detailed introduction to this
scheme will be given here, readers who are not familiar with it
may skip this section. Introductions
to the BV quantisation scheme can be found in \cite{me,L,Jordi,Henneaux}.

In the BV scheme one associates with every field $\phi^A$ an
antifield $\phi^*_A$ of opposite Grassmann parity. The antibracket
of two function(al)s $F$ and $G$ of the fields and antifields,
is defined by
\beq
   (F , G) = \frac{\dr F}{\delta \phi^A} \frac{\dl G}{\delta \phi^*_A} -
 \frac{\dr F}{\delta \phi^*_A} \frac{\dl G}{\delta \phi^A}.
\eeq
Canonical transformations within the antibracket
are generated by a Grassmann odd
function $F(\phi,\phi^{*'})$ of ghost\footnote{The ghost number of
the field and the ghost number of the
antifield are related by $\gh{\phi^A} + \gh{\phi^*_A} = -1$.}
number $-1$ .
Canonical transformation may produce a non-trivial Jacobian.

The classical action and the BRST transformation rules are brought
together in the {\it extended action S}. The most general
form needed for our present purposes is
\beq
           S = S_0(\phi) + \phi^*_A . \delta \phi^A \, .
\eeq
The extended action satisfies the classical Master Equation
$(S,S)=0$.

When considering a PV regularised theory, it is convenient to copy
the set-up one has for the fields and antifields to the PV fields,
$i.e.$:
\beq
   (F , G) = \frac{\dr F}{\delta \phi^A} \frac{\dl G}{\delta \phi^*_A}
   - \frac{\dr F}{\delta \phi^*_A} \frac{\dl G}{\delta \phi^A}
  + \frac{\dr F}{\delta \phi^A_{PV}} \frac{\dl G}{\delta \phi^*_{A,PV}} -
 \frac{\dr F}{\delta \phi^*_{A,PV}} \frac{\dl G}{\delta \phi^A_{PV}}.
\eeq

Our starting point is the regularised action of fermions coupled to an
external gauge field given in (\ref{start}).
One can imagine enlarging the degrees of freedom in the theory
with a scalar field $\theta$. Since the action $S_0$ does not depend
on $\theta$, there is a shift gauge symmetry $\delta\theta=\epsilon$.
Introducing a ghost field $c$ for this symmetry, one
obtains the extended action
\beq
\label{extact}
   S = \int d^2 x \left[\bar\psi i \dirac{D} \psi + \sum_l \bar\psi_l
   (i \dirac{D} - M_l) \psi_l + \theta^* c \right] \, ,
\eeq
which obviously satisfies the classical Master Equation.

The chiral rotation from the $\psi$ to the $\chi$ variables can be
implemented by a canonical transformation generated by
\beq
  F_1 = \theta^{*'}\theta + c^{*'}c + \bar \psi e^{-i\gv\theta} \bar
  \chi^* + \chi^* e^{-i\gv\theta}
   \psi + \bar \psi_l e^{-i\gv\theta} \bar \chi^*_l + \chi^*_l
   e^{-i\gv\theta} \psi_l  \, .
\label{Fone}
\eeq
This generating fermion produces the following transformations of the
fields, antifields and their regularisation counterparts (we suppress
spinor indices):
\begin{equation}
\begin{array}{cc}
   \chi = \frac{\delta F_1}{\delta\chi^*} =  e^{-i\gv\theta} \psi &
\chi_l = \frac{\delta F_1}{\delta\chi^*_l} =  e^{-i\gv\theta} \psi_l  \\
\\
     \bar \chi = \frac{\delta F_1}{\delta \bar \chi^*} = \bar  \psi
e^{-i\gv\theta}   &
 \bar \chi _l = \frac{\delta F_1}{\delta \bar \chi^*_l } = \bar  \psi_l
e^{-i\gv\theta}   \\
\\
  \theta' = \frac{\delta F_1}{\delta \theta^{*'}} = \theta  & c' =
\frac{\delta F_1}{\delta c^{*'}} = c \, ,
\end{array}
\end{equation}
which corresponds to the transformation (\ref{trans}) and (\ref{transreg}).

However, (\ref{Fone}) also determines the transformation of the
antifields. In particular, one finds
\beq
     \theta^* =  \frac{\delta F_1}{\delta \theta} = \theta^{*'}
     - i\bar \chi\gv \bar \chi^* - i \chi^* \gv \chi - i\bar \chi_l
     \gv \bar \chi^*_l  - i \chi^*_l \gv \chi_l  \, .
\eeq
Here the transformation rules of the fields have already been used.
The extended action (\ref{extact}) transforms into (dropping the primes
on $\theta$ and $c$)
\begin{eqnarray}\label{extact1}
  S_1 & = & \int d^2 x \left[\bar\chi i \Dt \chi + \sum_l \bar\chi_l (i
\Dt - M_l^\theta) \chi_l + \theta^* c \right. \nonumber \\
     & &  \left. - i\bar \chi\gv c \bar \chi^* + i \chi^* \gv c \chi -
i\bar \chi_l \gv c \bar \chi^*_l  + i \chi^*_l \gv c \chi_l  \right] \, .
\end{eqnarray}
This canonical transformation does not produce any Jacobian.

As was pointed out above, after the canonical transformation,
the classical action $S_1(\phi,\phi^*=0)$ has gauge symmetry (\ref{gsym1}).
Notice that by implementing the chiral rotation as a canonical
transformation, one immediately reads off the correct BRST
rules corresponding to this symmetry from the transformed extended
action (\ref{extact1}).

The next step is to make sure that the regulator fields get
the correct mass term $ - \bar \chi_l M \chi_l$ instead
of the mass term $ - \bar \chi_l M^\theta \chi_l$ that resulted from the
chiral rotation. Again this can be done using a
canonical transformation, where the appropriate generating fermion is
now
\beq
\label{Ftwo}
    F_2 = \unity  + \mu^*_l \frac{1}{i \Dt - M_l} (M_l - M_l^\theta)
    \chi_l \, .
\eeq
The only fields that change under this canonical transformation are
$\chi_l$,$\chi^*_l$ and $\theta^*$:
\beq
\label{tworules}
 \begin{array}{lll}
     \mu_l & = \frac{\delta F_2}{\delta \mu^*_l} & = \chi_l +
     \frac{1}{i \Dt - M_l} (M_l - M_l^\theta) \chi_l

      = \frac{1}{i \Dt - M_l} (i \Dt - M_l^\theta) \chi_l  \\
\\
     \chi^*_l  &  = \frac{\delta F_2}{\delta \chi_l} & = \mu^*_l +
     \mu^*_l \frac{1}{i \Dt - M_l} (M_l - M_l^\theta)

      =  \mu^*_l \frac{1}{i \Dt - M_l} (i \Dt - M_l^\theta) \\
  \\
     \theta^* & =  \frac{\delta F_2}{\delta \theta} & = \theta^{*'} +
     \frac{\dr }{\delta\theta} \left[
    \mu^*_l \frac{1}{i \Dt - M_l} (M_l - M_l^\theta) \chi_l
    \right] \, .
\end{array}
\eeq
The effect of this canonical transformation on the antifield independent
part of the extended action is trivial, it just produces the required
change from
\beq
    \bar \chi_l (i \Dt - M_l^\theta) \chi_l \longrightarrow
    \bar \chi_l (i \Dt - M_l) \mu_l  \, .
\eeq
The transformation on the antifield dependent part is somewhat more
complicated, and therefore we will provide some detail.
The two terms in (\ref{extact1}) that are going to change are
\beq
     \theta^* c + i \chi^*_l \gv c \chi_l       \label{twoterms}  \, .
\eeq
The second of these two terms (\ref{twoterms}) changes into
\beq
   T_1 = i \mu^*  \frac{1}{i \Dt - M} (i \Dt - M^\theta) .
   \gv c .  \frac{1}{i \Dt - M^\theta} (i \Dt - M)\mu \, ,
\eeq
where the subindex $l$ labelling the different PV copies is suppressed.
Let us now commute the $\gv c$ to the left until it is just behind the
antifield $\mu^*$.
The only trick to be used is the triviality
\beq
       [ {1 \over A} , B ] = - {1 \over A} [ A , B ] {1 \over A} \, .
\eeq
One then easily finds
\begin{eqnarray}
     T_1 & = & \mu^* i \gv c \mu \nonumber
         - \mu^* \frac{1}{i \Dt - M} \left( \gv \dirac{\partial} c
         + 2 \gv c \Dt \right) \mu \nonumber \\
    && + \mu^* \frac{1}{i \Dt - M} \left( \gv \dirac{\partial} c
    + 2 \gv c \Dt \right)  \frac{1}{i \Dt - M^\theta}
    (i \Dt - M ) \mu \,  .
\end{eqnarray}
To perform the canonical transformation on the first term in
(\ref{twoterms}) one has to be slightly more careful. It transforms to
\beq
   T_2 = \theta^* c + \int d^2 y \int d^2x \,\, \mu^*(x)
   \frac{\dr }{\delta\theta(y)} \left[
     \frac{1}{i \Dt^x - M} (i \Dt^x - M_l^{\theta(x)})  \right]
     . \frac{1}{i \Dt^x - M^{\theta(x)}} ( i \Dt^x - M ) \mu(x) c(y) \, .
\eeq
Again the index $l$ was suppressed and we have made all hidden
space-time coordinates  explicit. Probably the quickest way to
proceed is to use that  $\partial^x_\mu\delta(x-y) = -
\partial^y_\mu\delta(x-y)$. One finds
\begin{eqnarray}
 T_2 & = & \theta^*c + \mu^* \frac{1}{i \Dt - M} 2i \gv c M \mu
  + \mu^* \frac{1}{i \Dt - M} \left( \gv \dirac{\partial} c
  + 2 \gv c \Dt \right) \mu \nonumber \\
    && - \mu^* \frac{1}{i \Dt - M} \left( \gv \dirac{\partial} c
    + 2 \gv c \Dt \right)  \frac{1}{i \Dt - M^\theta} (i \Dt - M )
    \mu \,  .
\end{eqnarray}
Several terms cancel between $T_1$ and $T_2$, and one
obtains
\beq
   \theta^* c + i \chi^*_l \gv c \chi_l
   \stackrel{\mbox{can. traf. 2}}{\longrightarrow}  \theta^* c
    + \chi^*_l \left( i \gv c + \frac{1}{i \Dt - M_l} 2i \gv c M_l
    \right) \chi_l \,
\eeq
where we have changed notation back to the $\chi$-fields.
Summarising, one finds that the extended action (\ref{extact1})
transforms into
\begin{eqnarray}\label{extact2}
    S_2 & = & \int d^2 x \left[\bar\chi i \Dt \chi + \sum_l \bar\chi_l
    (i \Dt - M_l) \chi_l + \theta^* c \right. \nonumber \\
     & &  \left. - i\bar \chi\gv c \bar \chi^* + i \chi^* \gv c \chi
     - i\bar \chi_l \gv c \bar \chi^*_l  + \chi^*_l \left(i \gv c
     +  \frac{1}{i \Dt - M_l} 2i \gv c M_l \right)    \chi_l  \right]
     \, .
\end{eqnarray}
The canonical transformation generated by (\ref{Ftwo}) has
also produced
a Jacobian. As is obvious from (\ref{tworules}), it is precisely the
one shown in (\ref{stap2}).

We now construct a so-called non-minimal extended action,
$
    S_2^{n.m.} = S_2 + \bar c^* b ~,
$
and perform a third canonical transformation generated by
\beq
     \label{Fthree}
      F_3 = \unity + \Psi_R \, .
\eeq
Since $ \Psi_R$ only depends on fields, only the antifields will
transform non-trivially under this canonical transformation, and
therefore the Jacobian is $1$. The extended action changes precisely
by the addition of $\delta\Psi_R$ of (16):
\beq
       S_3  = S_2^{n.m.} + \delta\Psi_R \, .
\eeq
At this point, one can put the antifields to zero, and go through all
the steps described above. In this way we have rederived bosonization
in the BV scheme. Introducing a free parameter $\Delta$ as explained
below, we also rederive the generalization, smooth bosonization, in
BV language.

{\bf 4.}
The procedure we have outlined above is of course general, and in principle
straightforward. The most important aspect to keep in mind is that
the regulator fields must be kept explicitly in the path integral
throughout, and in particular when one is gauge-fixing (or changing
gauges) by the addition of BRST-exact terms. The second requirement
is the natural one of always manipulating properly regularized expressions.

In ref. \cite{us3} a more convenient way of achieving the same kind
of ``quantum mechanical gauge" that we have derived above was described.
It hinges rather crucially on the situation peculiar to the bosonization
example discussed here: the collective field $\theta(x)$ which
is responsible for the new gauge symmetry does not require additional
regulator fields.\footnote{Intuitively this is easy to understand, since
the whole path integral is regularized to begin with, and redefinitions
of the fields do not affect this overall regularization. This holds
even if we integrate over the additional collective fields, because the
gauge-fixing is precisely removing the new redundant propagating degree
of freedom.} However, some aspects have more general validity, and
since it very nicely elucidates the issue of gauge fixing in
the presence of regularization, we shall now outline it.

Let us first state in more clear terms which gauge we finally wish to
achieve. In the bosonization gauge of smooth bosonization \cite{us1},
the idea is
to find a gauge such that the physical (and gauge invariant)
axial current $\psi\gamma^{\mu}\gamma_5\psi$
is described entirely by a bosonic expression. This is set up to
hold in the full generating functional (and of course then also off
mass shell).

As we have shown above, this is difficult (but possible) because it
involves gauge-fixing away the chirally rotated fermionic axial current
$\partial_{\mu}\bar{\chi}\gamma^{\mu}\gamma_5\chi$, and in its
unregularized form this object is gauge (or BRST) invariant. One way
to cure the problem is, as we have demonstrated above, to keep explicitly
the regulator fields in the path integral. But suppose we adhere to
the formulation in which the regulator fields have already been integrated
out, leaving a non-trivial regularized $\bar{\chi}, \chi$-measure:
how do we proceed?
We shall now show that a short-cut exists in this formulation. The idea
is to first gauge-fix on an entirely unproblematic function, and then
shift the axial source in a BRST invariant manner so as to finally
reach the non-trivial gauge we seek. That such a procedure is possible
is due to the fact that the gauge symmetry we are considering arises
from a field redefinition (a chiral rotation) of exactly the same kind
as the gauge transformation itself.
The Jacobian we obtain from the field redefinition is then in form
completely equivalent to the Jacobian we subsequently get under BRST
transformations.

When only the $\bar{\psi}, \psi$-measure
contains the regularization, already the first step (eq. (\ref{trans}))
above is modified. The Jacobian of the field redefinition is now
non-trivial, and the analogue of eq. (\ref{Zext}) therefore contains a
quantum correction:
\begin{eqnarray}
Z_{ext} &=& \int d[\theta]d[\bar{\chi},\chi]_\Lambda \exp\left(\frac{i}
{\hbar}S_{ext}\right) \cr
S_{ext} &=& \int d^2x\left[\bar{\chi}i\slash{D}_{\theta}\chi
+i\hbar {\cal M}_1^{(\Lambda)}(\theta,A_{\mu})\right]
\end{eqnarray}
Here, the Jacobian ${\cal M}_1^{(\Lambda)}$ can be expressed in terms of
an infinite expansion in inverse powers of the ultraviolet
cut-off $\Lambda$:
\bea
{\cal M}_1(\Lambda ) &=&  -i \ln \det \prod_l
\left( \frac{i\slash{D}_\theta
-M_l}{i\slash{D}_\theta -M_l^\theta} \right)^{c_l}\cr
&=& \int\! d^2x\; \left[ \frac{1}{2\pi} \partial_\mu \theta
\partial^\mu \theta -\frac{1}{\pi} A_\mu \partial^\mu \theta
+ \frac{\kappa_{-2}}{12 \pi \Lambda^2} \partial^2 \theta \partial^2
\theta - \frac{\kappa_{-2}}{6 \pi \Lambda^2} \partial_\nu A_\mu
\partial_\nu \partial_\mu \theta + {\cal O} (\Lambda^{-4})\right]
\eea
The cut-off itself is related to the Pauli-Villars regulator masses
as defined in (\ref{PVdef}).

The idea of ref. \cite{us3} is now to first gauge-fix in the more
simple $\theta$-sector. This may sound paradoxical, since in the
bosonization gauge we precisely wish to retain fully the
$\theta$-field. However, the next step consists in shifting the
external source $A_{\mu}$ by the derivative of the
Lagrangian multiplier,  $\partial_{\mu}b$. As we
will see, this can be done in such a way as to precisely cancel
the effect of gauge fixing on $\theta$. We thus first add
\beq
-\frac{i\hbar}{\pi} \delta(\bar{c}\partial^2\theta) = -\frac{i\hbar}{\pi}
b\partial^2\theta + \frac{i\hbar}{\pi} \bar{c}
\partial^2 c
\eeq
to the action.\footnote{Note the explicit factor of $\hbar$ here.}
This is a correct gauge fixing term for setting
$\partial^2\theta = 0$. Next, we shift the source: $A_{\mu} \to
A_{\mu} + \partial_{\mu}b$. This is a BRST-invariant procedure because
$A_{\mu}$ couples only to a term that has vanishing BRST-variation, and
$b$ itself is neutral under BRST transformations. The result is an
action of the form
\begin{eqnarray}
 S_{ext}  &= &\int d^2x\left[\bar{\chi}i\slash{D}_{\theta+b}\chi
+ i\hbar {\cal M}_1(A_{\mu} \to A_{\mu} + \partial_{\mu}b) -
\frac{i\hbar}{\pi} b\partial^2\theta + \frac{i\hbar}{\pi}
\bar{c}\partial^2 c \right]  \cr
&=& \int d^2x\left[\bar{\chi}i\slash{D}_{\theta+b}\chi
+ i\hbar\ldots  + \frac{i\hbar}{\pi} \bar{c}\partial^2 c \right]
\eea
Note that there is a term $b\partial^2\theta$ from ${\cal M}_1$ (see
(\ref{M1})) which just cancels the gauge fixing term we started with.
With this modification,  we precisely identify
\beq
\partial_{\mu}J^{\mu}_5 ~\sim~ -1/\pi \partial^2 \theta ~.
\eeq
This can be seen easily by shifting $b$ by $\varphi$ with $A_\mu =
\partial_\mu \varphi$. This shift removes all occurrences of the
external source $\varphi$ except for its coupling to $\partial^2\theta$.

A rephrasing of this procedure in the limit of infinite
cutoff has recently been discussed in ref. \cite{Theron}. Duality
transformations between bosons and fermions in (1+1) dimensions also
effectively rely on gauge fixings in sectors that are free of
complications due to regularization \cite{Burgess}.

The biggest advantage of the procedure described above
is that in general the precise content of the gauge fixed theory
is completely known. It gives an exact specification of the
generating functional in terms of the degrees of freedom we choose,
and hence all relevant Green functions computed in terms of those
degrees of freedom. Let us finally formulate the procedure in a more
general manner.

We start with a functional integral governed by an action $S$
parametrized as follows:
\beq
S[\phi^A,A_r] + \hbar {\cal M}_1 [\phi^A,A_r]
\eeq
where $\phi^A$ and $A_r$ denote the various fields and external
sources. The theory has a gauge symmetry
\beq
\delta\phi^A = R^A_i[\phi]\epsilon^i
\eeq
such that the Jacobian of this transformation is cancelled
by the variation of $M_1$. The sources $A_r$ are chosen to couple only
to gauge invariant operators ${\cal O}_r$. The generating functional is
thus gauge invariant even for non-vanishing sources.

Suppose that a subset of the fields $\phi^A$, denoted by
$\phi^i$, transforms under this gauge transformation with unit
Jacobian. Furthermore, suppose we wish to find the gauge that
identifies these fields $\phi^i$ with operators ${\cal O}^i$
that couple to the sources $A_i$; this, of course, implies that there
are couplings $A_i \phi^i$. We can achieve this by first
gauge fixing the fields $\phi^i$ to zero by adding
\beq
\delta \Psi = \delta \left( \bar{c}_i \phi^i \right) = b_i \phi^i -
\bar{c}_i R^i_j c^j ~.
\eeq
Next, shift the source $A_i \to A_i - b_i$, and subsequently
shift $b_i \to b_i + A_i$. The source $A_i$ then couples only to $\phi^i$,
$i.e.$, we have reached the gauge in which the (gauge invariant)
operator ${\cal O}^i$
is represented by the (gauge dependent) field $\phi^i$.
A gauge aiming at identifying  $\phi^i$
with only a given fraction $\Delta$ of the operator
${\cal O}^i $ can be obtained by replacing the shift of $A_i$ above by
$A_i \to A_i - \Delta b_i$. In smooth bosonization, $\Delta$ is the
parameter by means of which one can interpolate smoothly between a
bosonic and a fermionic gauge.
\vspace{0.5cm}
\noindent

{\sc Acknowledgment:} The work of FDJ was supported by the Human Capital
and Mobility Programme through a network on Constrained Dynamical
Systems.

\end{document}